# Comparison of soluble and insoluble organic matter in analogues of Titan's aerosols

Julien MAILLARD[1, 2,*], Nathalie CARRASCO[1], Isabelle SCHMITZ-AFONSO[2], Thomas GAUTIER[1], Carlos AFONSO[2]

[1] LATMOS/IPSL, Université Versailles St Quentin, UPMC Université Paris 06, CNRS, 11 blvd d'Alembert, F-78280 Guyancourt, France

[2] Université de Rouen, Laboratoire COBRA UMR 6014 & FR 3038, IRCOF, 1 Rue Tesnière, 76821 Mont St Aignan Cedex, France

***Corresponding author:**
Julien MAILLARD
julien.maillard@ens.uvsq.fr
1 Rue Lucien Tesnière, 76130 Mont-Saint-Aignan, France

**Abstract**

Titan, the biggest moon of Saturn, has a thick atmosphere which presents similarities with the one thought to be on Earth at its beginning. The study of Titan's photochemical haze is thus a precious tool in gaining knowledge of the primitive atmosphere of Earth. The chemistry occurring in Titan's atmosphere and the exact processes at act in the formation of the hazes remain largely unknown. The production of analogs samples on Earth has proved to be a useful tool to improve our knowledge of the aerosols formation on Titan. Such solid organic analogs samples, named tholins, were produced with the PAMPRE experiment (French acronym for Aerosols Microgravity Production by Reactive Plasma). PAMPRE tholins were found to be mostly insoluble, with only one-third of the bulk sample that can be dissolved in methanol. This partial solubility limited the previous studies in mass spectrometry, which were done only on the soluble fraction. The goal of the present study is to compare the two fractions of PAMPRE's tholins (insoluble and soluble) using a ultra-high resolution Fourier transform ion cyclotron resonance mass spectrometer (FTICR) equipped with a laser desorption/ionization source. Using modified Van Krevelen diagrams, we compare the global distribution of the molecules within the samples according to their Hydrogen/Carbon ratio and Nitrogen/Carbon ratio. Major differences are observed in the molecular composition of the soluble and the insoluble fraction. The soluble fraction of tholins was previously identified as a set of polymers of average formula $(C_2H_3N)_n$. In this work we observe that the insoluble fraction of tholins is comprised of a significantly different set of polymers with an average composition of $(C_4H_3N_2)_n$.

## Graphical abstract

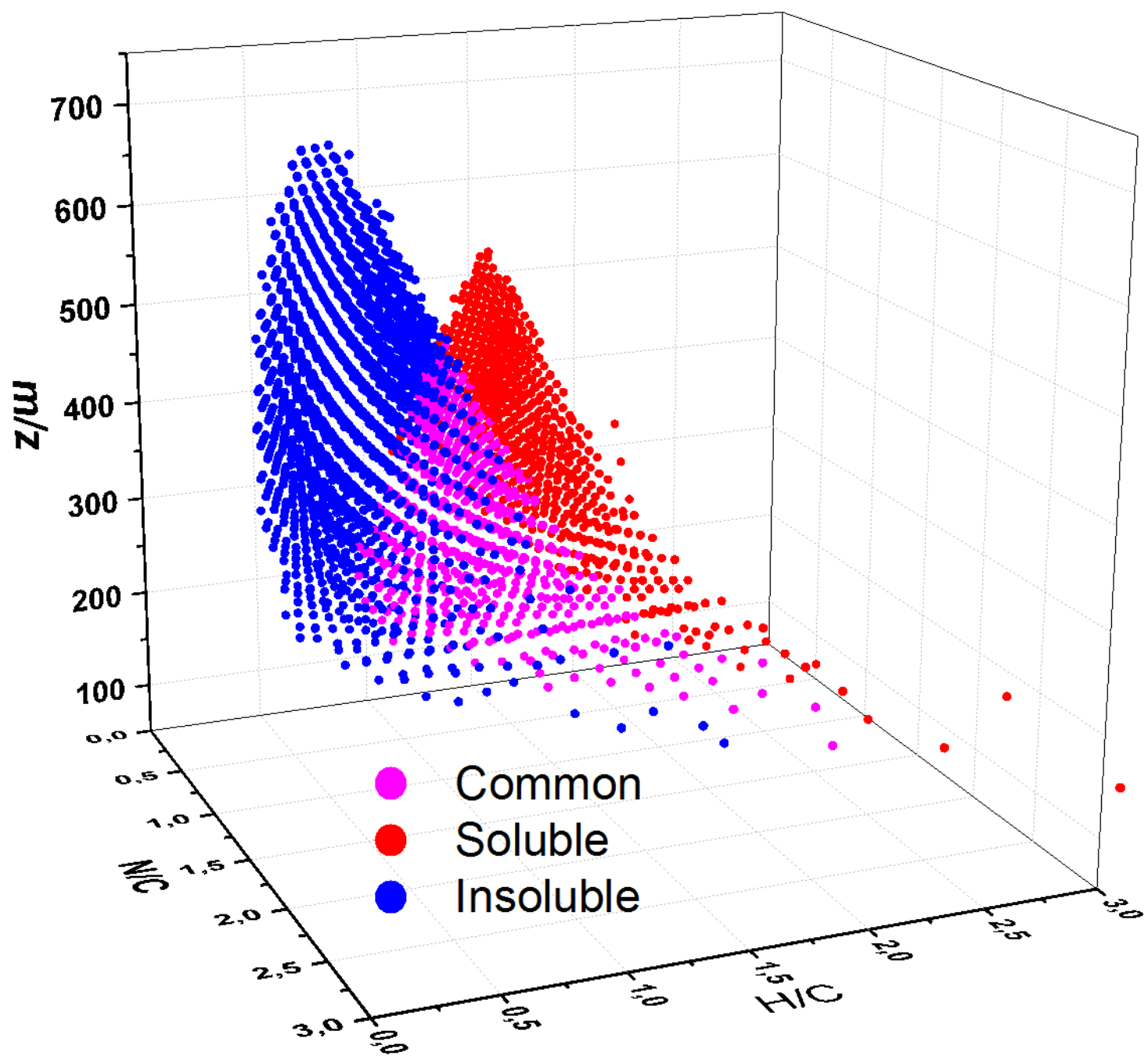

## Introduction

Titan, the largest satellite of Saturn, is surrounded by a thick atmosphere rich in nitrogen with a small percentage of methane. The chemistry in this atmosphere leads to the formation of a complex smog due to the radicalization and recombination of the two gases upon interaction with solar UV photons and charged particles from Saturn's magnetosphere. The observations of the Cassini mission and the information obtained during the landing of the Huygens probe provided partial information about the composition of the aerosols. Pyrolysis of the aerosols by the ACP experiment aboard Huygens released mainly $NH_3$ and HCN but was not sufficient to further elaborate on the structural composition of the aerosols because of the detection limit of the instrument (Fulchignoni et al., 2005; Israel et al., 2005; Niemann et al., 2005).

Titan's aerosols also represent an interest for astrobiology (Sagan et al., 2002) due to their peculiar formation and their organic complexity. Some analogs, called tholins, were produced on Earth to investigate the formation processes and the structure of Titan's aerosols (Cable et al., 2012; Imanaka and Smith, 2010; Somogyi et al., 2005; Szopa et al., 2006; Toupance et al., 1975).

Infrared (IR) spectroscopy analyses have demonstrated that tholins structure includes nitrile, amine, hydrocarbon and unsaturated functions (Cable et al., 2014; Coll et al., 1999; Gautier et al., 2012; Imanaka et al., 2004; Sciamma-O'Brien et al., 2017). Primary amines were also identified in tholins using laser-induced fluorescence with nonaqueous capillary electrophoresis (Cable et al., 2014). Additional analyses were performed on the soluble fraction of the tholins by mass spectrometry showing a polymeric structure (Bonnet et al., 2013; Gautier et al., 2016; Vuitton et al., 2010). First analyses with ultra-high resolution mass spectrometry also brought to light repetition patterns ($CH_2$ and HCN) in the tholins' soluble fraction (Anicich et al., 2006; He et al., 2012; Somogyi et al., 2005).

However, some tholins are not completely soluble in solvents such as methanol (Carrasco et al., 2009). This is the case of the tholins synthesized with the PAMPRE experiment, described below (Szopa et al., 2006). The soluble fraction of tholins obtained from this experiment represents only one third of the total material. Only this soluble fraction was analyzed with mass spectrometry so far, using electrospray ionization and photoionization (Carrasco et al., 2009). The soluble fraction provided numerous information about the general structure of these samples. Several studies compared various tholins to a set of polymeric structures based on a $CH_2$-HCN repeating unit, spreading from *m/z* 100 to *m/z* 800, as emphasized by the Kendrick mass defect diagram using HCN or $CH_2$ normalization (Hughey et al., 2001; Kendrick, 1963). This explained the periodic nature of the mass spectra of tholins showing repeating ion clusters (Gautier et al. 2014, Pernot et al. 2010).

A previous study emphasized that IR spectrum of the soluble and non-soluble fractions of tholins were nearly identical (Carrasco et al., 2009). As a consequence, it has been postulated that both fractions had a similar molecular content, and that the difference of solubility was only related to the difference of mass of the molecules composing tholins, as larger molecules are expected to be less soluble. In the present study, we reexamine this insoluble fraction at the molecular level by using ultra-high resolution mass spectrometry.

## Experimental

### *Sample production*

Tholins were produced using the PAMPRE experiment and following the procedure detailed in previous publications (Gautier et al., 2011; Szopa et al., 2006). In this setup, the reactor is composed of a stainless steel cylindrical reactor in which the Radio Frequency – Capacity Coupled Plasma (RF-CCP) discharge is established by an RF 13.56 MHz frequency generator. The gas mixture, containing nitrogen and 5% of methane, is injected in the chamber as a continuous flow through polarized electrodes and then extracted by a primary vacuum pump to ensure that gases are homogeneously distributed. The plasma discharge is maintained at a pressure of $0.9 \pm 0.1$ mbar and at room temperature.

### *Sample treatment for mass spectrometry analysis*

In order to separate soluble and insoluble fractions, 4 mg of tholins were dissolved in 1 mL of methanol in a vial under ambient atmosphere. The vial was vigorously stirred for 3 minutes to dissolve the maximum amount of species. The brown mixture was then filtered using a 0.2 μm polytetrafluoroethylene (PTFE) membrane filter on a filter holder. Of the filtered solution, the soluble fraction, was transferred in a glass vial. Half dilution with a 50/50 water/methanol mixture was performed just before analysis in order to be analyzed using similar conditions as previous studies by electrospray ionization. The PTFE membrane was then recovered, placed in a vial and left open under a neutral atmosphere of Nitrogen to evaporate the remaining methanol and avoid contamination. The insoluble fraction, recovered as a black powder from the membrane, was then analyzed by mass spectrometry. The global sample was directly analyzed avoiding any contact with solvents.

*Mass spectrometry analyses*

Different ionization methods can be used for the analysis of molecules in their solid state, especially Laser Desorption Ionization (LDI). LDI-MS has already been used for the analysis of tholins (Gautier et al., 2017; Imanaka et al., 2004; Mahjoub et al., 2016; Sagan et al., 1993; Somogyi et al., 2012). It has also been used for the analysis of asphaltenes, a fraction of petroleum non-soluble in pentane or heptane (Pereira et al., 2014; Sagan et al., 1993; Tanaka et al., 2004). In this work a LDI source is coupled with a Fourier transform ion cyclotron resonance mass spectrometer (FTICR) to compare the soluble and non-soluble fractions of tholins at the molecular level. A comparison between electrospray and LDI sources using the same analyzer parameters can be found in the supporting information (Figures S1 and S2). In addition, LDI-FTICR was compared to LDI-TOF analyses (Figure S1).

Previous works have demonstrated the significant differences in the molecules detected in tholins depending on the ionization mode used (Carrasco et al., 2009; Somogyi et al., 2012). For this work we chose to investigate only positive ionization as this is the one most widely used in the community for tholins analysis.

All analyses were performed on a FTICR Solarix XR from Bruker equipped with a 12 Tesla superconducting magnet and a laser desorption ionization source (laser NdYAg 355 nm). The mass spectrometer was externally calibrated with a solution of sodium trifluoroacetate. Mass spectra were afterwards internally calibrated with confidently assigned signals yielding a mass accuracy bellow 300 ppb in the considered mass range. The soluble fraction was deposited on a LDI plate using the dry droplet method (5 x 1 μL to ensure a good concentration of the sample). Insoluble fraction and the global sample were deposited using a solvent-free method, following a previously published procedure (Barrere et al., 2012). Mass spectra were recorded in positive mode at 8 million points with a sum of 500 scans, yielding a resolution of 1 500 000 at *m/z* 150 and 500 000 at *m/z* 500. The following instrumental parameters were implemented for the insoluble and total fractions: Plate offset 100 V, Deflector plate 210 V, Laser power 19 %, Laser shots 40, Frequency of laser shots 1000 Hz, Funnel 1 at 150 V, Skimmer 1 at 25 V. The same parameters were used for the soluble fraction, except the laser power that was raised to 29% and the number of shots increased to 150. Fraction responses to the laser ionization were different for each fraction of the sample due to the molecular differences between each fraction, especially the hydrogen/carbon ratio. To obtain sufficient signal for the soluble fraction, it was necessary to increase the laser power compared to the parameter used for the non-soluble fraction spectra. A comparison between the global intensity of the spectra and the laser power used for the ionization is given in the supplementary material (Figure S3). We then defined a laser power value for each fraction in order to obtain similar signal intensities. We fixed the value at 5% of laser power above the determined threshold – 19 % and 29% for the insoluble and soluble fraction respectively. Peak picking was done with a signal/noise ratio of 5 and 0.01% of intensity. Molecular formulas were obtained using the SmartFormula tool from the Bruker Data Analysis 4.4 software with the following parameters: molecular formula $C_{0\text{-}x}H_{0\text{-}y}O_{0\text{-}2}$, even and odd electron configuration allowed, 0.5 ppm error tolerance.

It should be noted that while we try to limit oxygen contamination, tholins are known to be prone to oxidation when exposed to oxygen. Numerous discussions can be found in the literature on this point (see for example (Carrasco et al., 2016) for a discussion on PAMPRE tholins). In our case, oxidation principally happens when tholins are exposed to the air during their collection after production. This exposition is intrinsic to the experimental setup used. Studies on plasma produced organic materials have shown that this oxidation happens within seconds of exposition of the sample to ambient air (Hörst et al., 2018; McKay, 1996; Swaraj et al., 2007; Tran et al., 2003). In the case of the sample analysed here, the amount of oxygen contamination can be estimated through the detection of oxygenated molecules in the samples from the LDI-FTICR measurement. These measurements show the presence of ~20% of oxygenated signals (Table 1). This part is non negligible principally because the material is composed of thousands of molecules in low abundance and so, thousands of molecules start to pick up oxygen after the exposition to ambient atmosphere. Nevertheless, especially because oxygenated species are produced from the bulk material and contain one or two oxygen atoms, no modification was observed between these oxygenated and non-oxygenated species concerning the number of unsaturation or the H/C ratio. The formation of such oxygenated species was observed previously upon the addition of ammonium hydroxide which were depicted as hydrolysis products (Neish et al., 2009; Poch et al., 2012). The oxygenated species were excluded from the results presented here by filtering the CSV data file obtained after molecular formulas attribution using Excel 2016 spreadsheet. Table 1 summarize all attributed formulas by the SmartFormula tool.

**Table 1 (1 column) : Classification of detected species by their families**

| | % | Attributed formulas |
|---|---|---|
| **Total ions** | **100** | **16952** |
| Non-oxygenated | 33 | 5538 |
| Oxygenated | 20 | 3472 |
| $^{13}C$ containing species | 21 | 3614 |
| $^{15}N$ containing species | 16 | 2652 |

## Results and discussions

### *LDI-FTICR analyses*

Figure 1 presents a general view of the recorded mass spectra for the two different fractions (soluble and insoluble) and for the global sample (soluble + insoluble). It can be seen that the base peaks are different in each fraction. For the soluble fraction, *m/z* 127.09784 is the base peak while it is *m/z* 200.09308 for the insoluble fraction and 214.10872 for the global sample. On the other hand, the ion distribution is relatively similar for all fractions but we observe that low *m/z* species are more abundant in the soluble fraction than in the other samples. The three mass spectra present a periodic profile with patterns consistent with the expected repeating unit of tholins $CH_2$ – HCN (Pernot et al., 2010).

The enlargement in the inset of Figure 1 highlights that for the insoluble fraction and for the global sample, the formed repetition pattern are similar, while they appear shifted in the soluble fraction. This phenomenon is better illustrated on Figure 2 by zooming in on several periods. We can observe that the soluble fraction has a period (two repetition patterns) of *m/z* 27.0109, which corresponds exactly to the HCN formula. In comparison, insoluble and the global sample have a period of *m/z* 26.0156, which corresponds exactly to the $C_2H_2$ formula. This explains why in some parts of the spectra, the waves of the soluble fraction present their lowest intensity and their highest intensity for the two different fractions. These results highlight that significant differences exist between the soluble and non-soluble samples in the growth structure of the tholins. However, these two specific periods are not the only one detected in each fraction. In addition to this $C_2H_2$ growth pattern, there are actually numerous ions series present in the samples. Previous work from Gautier et al. 2014 showed that one would need at least 8 different growth series to represent the samples, many of them bearing nitrogen thus explaining the presence of N-containing compounds at relatively high molecular masses. We consider that presenting a full description for all ion series goes beyond the scope of the present paper and chose to focus solely the $C_2H_2$ as an illustration of the phenomenon.

The obtained results are also consistent with those recorded in previous studies on tholins produced on another experimental setup according to the spectrum appearance (He et al., 2012; Somogyi et al., 2005).

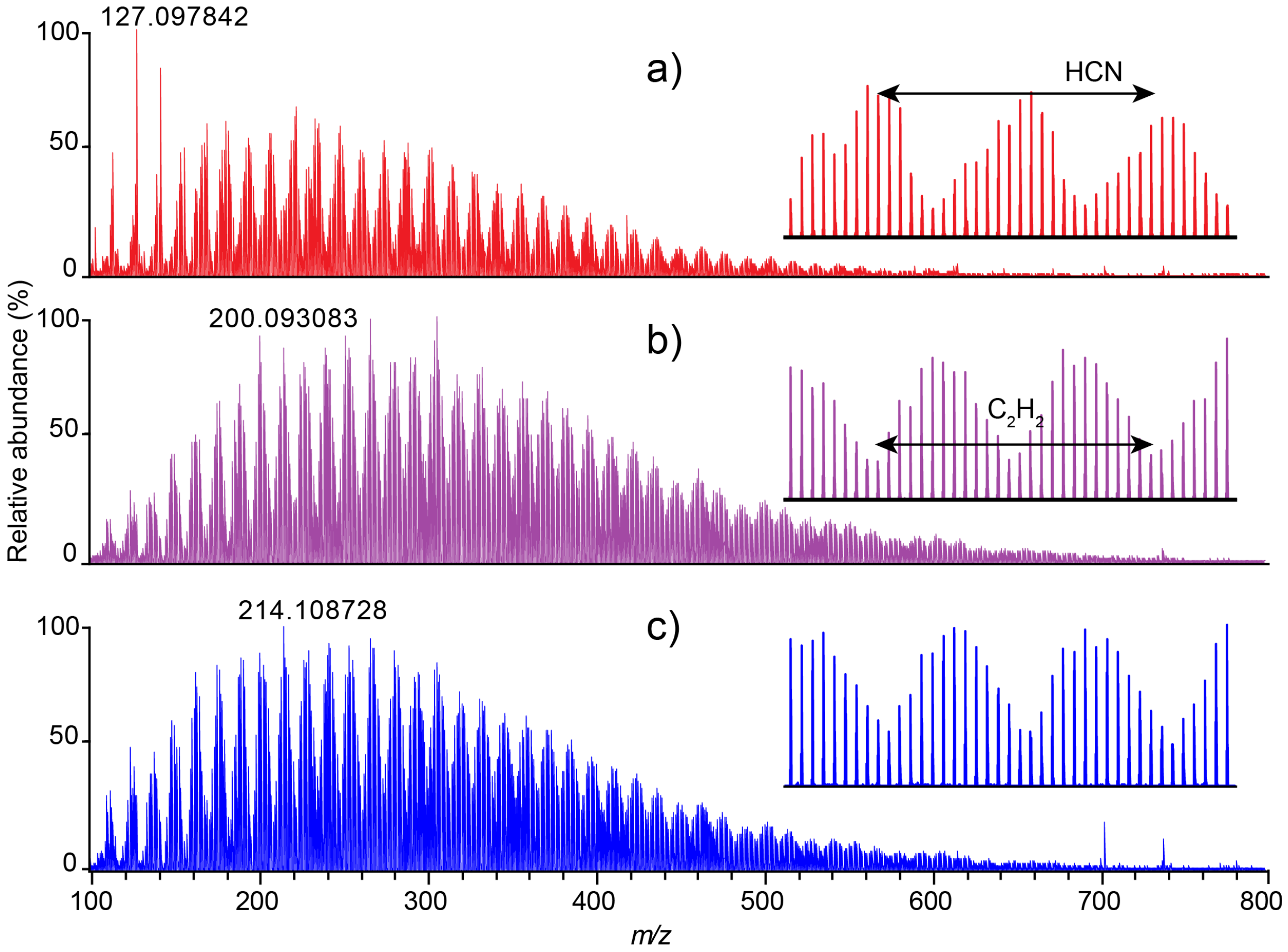

Figure 1 (2 columns): Laser desorption/ionization FTICR MS spectra of a) soluble fraction, b) non-soluble fraction, c) global sample

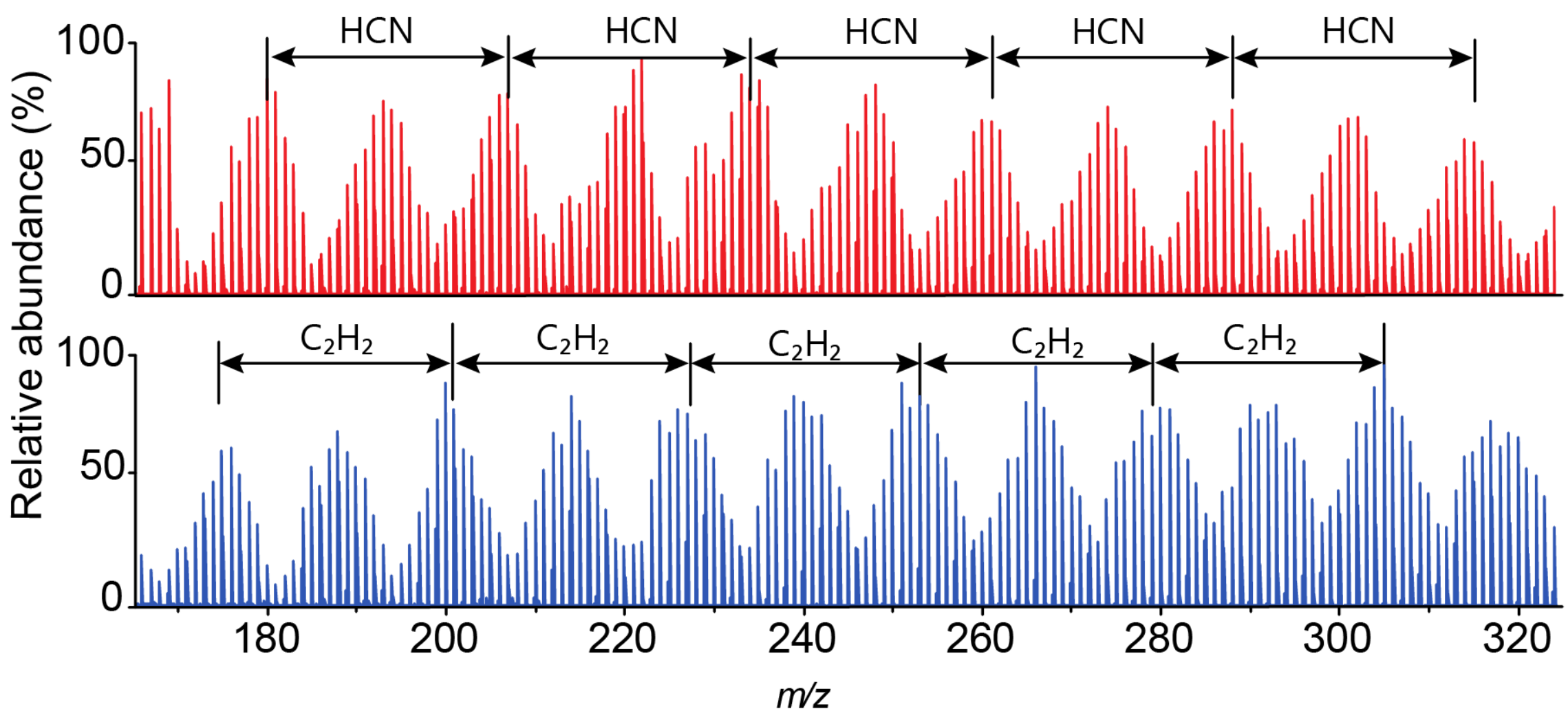

Figure 2 (2 columns): Enlargement on the soluble (in red) and non-soluble fraction (in blue) to highlight the difference between the periodic phenomenon on each fraction

Figure 3 shows different sections of the mass spectra centered at *m/z* 148 and *m/z* 512, for the two different fraction and the global sample. These two parts of the spectrum were chosen to highlight the differences for the low and high *m/z* values. The ultra-high resolution of the FTICR instrument allows the separation of the isobaric ions even for highly complex ion distributions evidenced at the higher mass range. Interestingly, the three spectra are almost identical in the low mass range but differences can be observed for the higher mass range, with the detection of ions present in the soluble fraction but not in non-soluble fraction. As one can see on the enlargement of Figure 3 on *m/z* 512, the signal for the global sample seems to correspond to the sum of the signals present in the soluble and non-soluble fraction. The obtained ion relative abundance shows a higher amount of species specific to the non-soluble fraction in the total fraction. This is consistent with the amount of soluble species, which corresponds to only 35% of the total tholins (Carrasco et al., 2009).

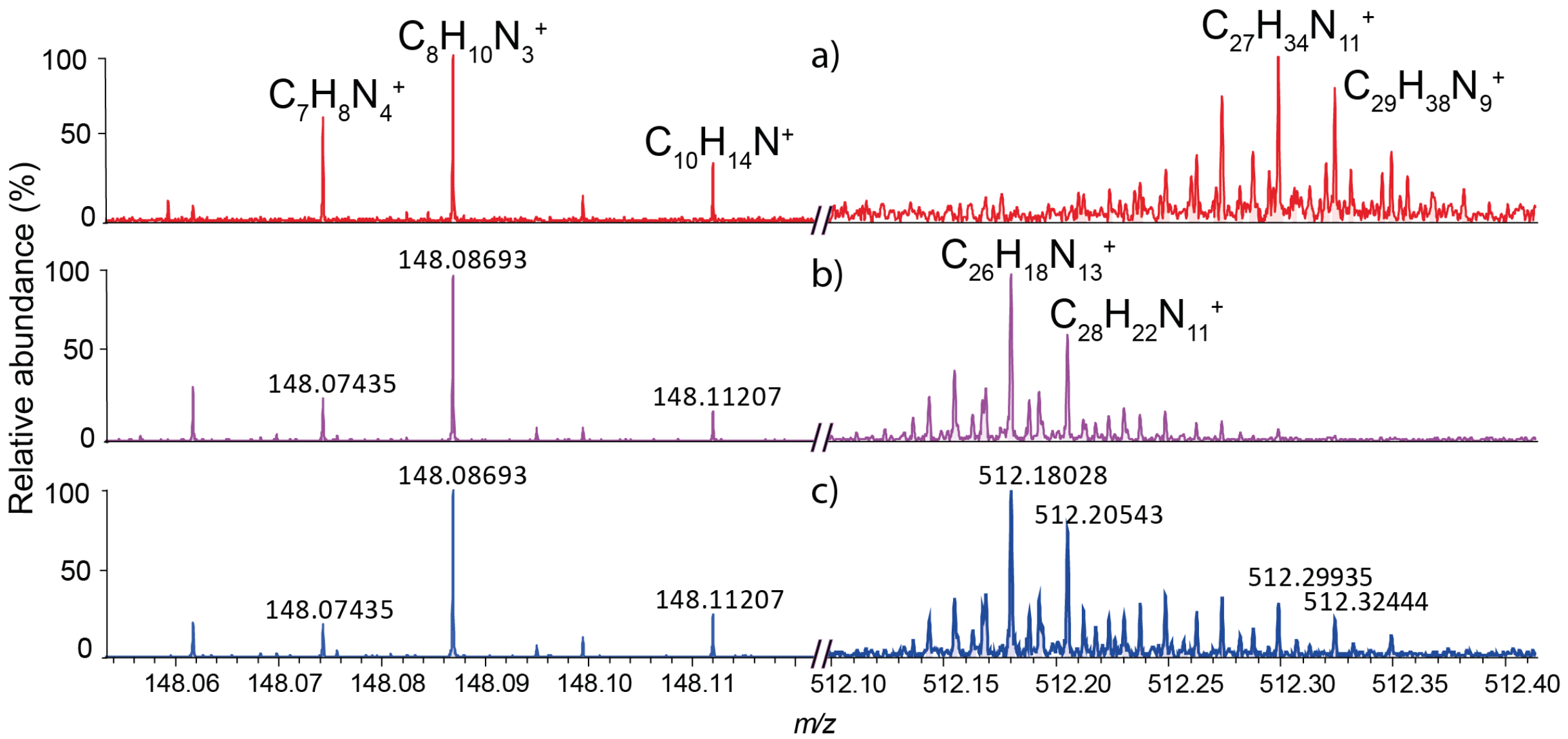

Figure 3 (2 columns): Enlargement of the *m/z* 148 and *m/z* 512 ions for a) soluble fraction, b) non-soluble fraction, c) global sample

In order to obtain a broader picture on the detected species, molecular formulas were determined for each ion. This was done by considering only species presenting C, H and N elements. Due to the mass accuracy of our measurement (<300 ppb), only one molecular formula could fit with each experimental mass. Looking into details at the molecular formula (see for example *m/z* 512 on Figure 3), the ions specifically detected in the global sample and non-soluble fractions present a low mass defect, consistent with less hydrogenated species. This comparison emphasizes once again important differences between the soluble and the non-soluble fractions: not on the mass of the molecules as initially thought, but rather in the type of molecules present in each sample.

*Data treatment using modified Van Krevelen diagrams*

Van Krevelen diagrams were originally proposed for displaying the molecular complexity of petroleum or coal with bidimensional graphics (Kim et al., 2003; Van Krevelen, 1950). Such diagrams were used more recently for other research fields including global 'omics' analysis of metabolites but also tholins (Pernot et al., 2010). In typical Van Krevelen diagrams O/C are plotted as a function of H/C ratio. These ratios are readily obtained from the molecular formulas. These diagrams particularly show evidence of unsaturation based on the H/C ratio and the number of hetero-elements with the O/C ratio. Previous studies in tholinomics demonstrated the interest of these diagrams by switching the x-axis that commonly represents the oxygen/carbon ratio with the nitrogen/carbon ratio (Gautier et al., 2014; Imanaka and Smith, 2010; Somogyi et al., 2005). Van Krevelen diagrams are limited because they do not consider the *m/z* dimension and ions with very different m/z ratio can yield the same H/C or N/C ratio. The *m/z* dimension can be added to the Van Krevelen plot to generate 3D plots (Gautier et al., 2014; Pernot et al., 2010). Specific 2D plots H/C vs m/z or N/C vs m/z have also been used (Imanaka and Smith, 2010).

A comparison was done to choose the most appropriate axis to be used to represent our samples. Soluble and insoluble fractions were found quite equal in their nitrogen/carbon ratio as visible in the Figure 4-right. The nitrogen/carbon ratio bearing little information, the most interesting representation for our samples was determined to be the *m/z* value versus the ratio hydrogen/carbon (Figure 4 left). 3D Van Krevelen to highlight the interest for the first global view on the sample based on the data presented in the Figure 4 can be found in supporting information (Figure S2).

Figure 4 shows the H/C versus *m/z* plot and the N/C versus *m/z* plots obtained from the soluble (in red) and the insoluble fractions (in blue). We focused our study on the major contributors to the sample. Thus only species with an intensity above 5% of the base peak (corresponding to 81% of all species detected in the samples according to the sum of all peaks intensities) are displayed. In this illustration, we observe for the soluble fraction (including the common part between the different samples in purple) a triangular distribution that converges toward a unique peak. This distribution has been observed in a previous work (Pernot et al., 2010). This structure starts at a low *m/z* value with a large base and then becomes thinner with the increase of the *m/z* value. This shape is in agreement with a set of copolymers involving the –$CH_2$-HCN- repeating unit with an addition of some carbon in the structure that enlarges the distribution at low *m/z* values. This confirms the previous results on the soluble fraction described in Pernot et al., 2010, a previous study realized with an Orbitrap$^{TM}$ analyzer with a resolution of 200 000 at *m/z* 150 and 100 000 at *m/z* 450.

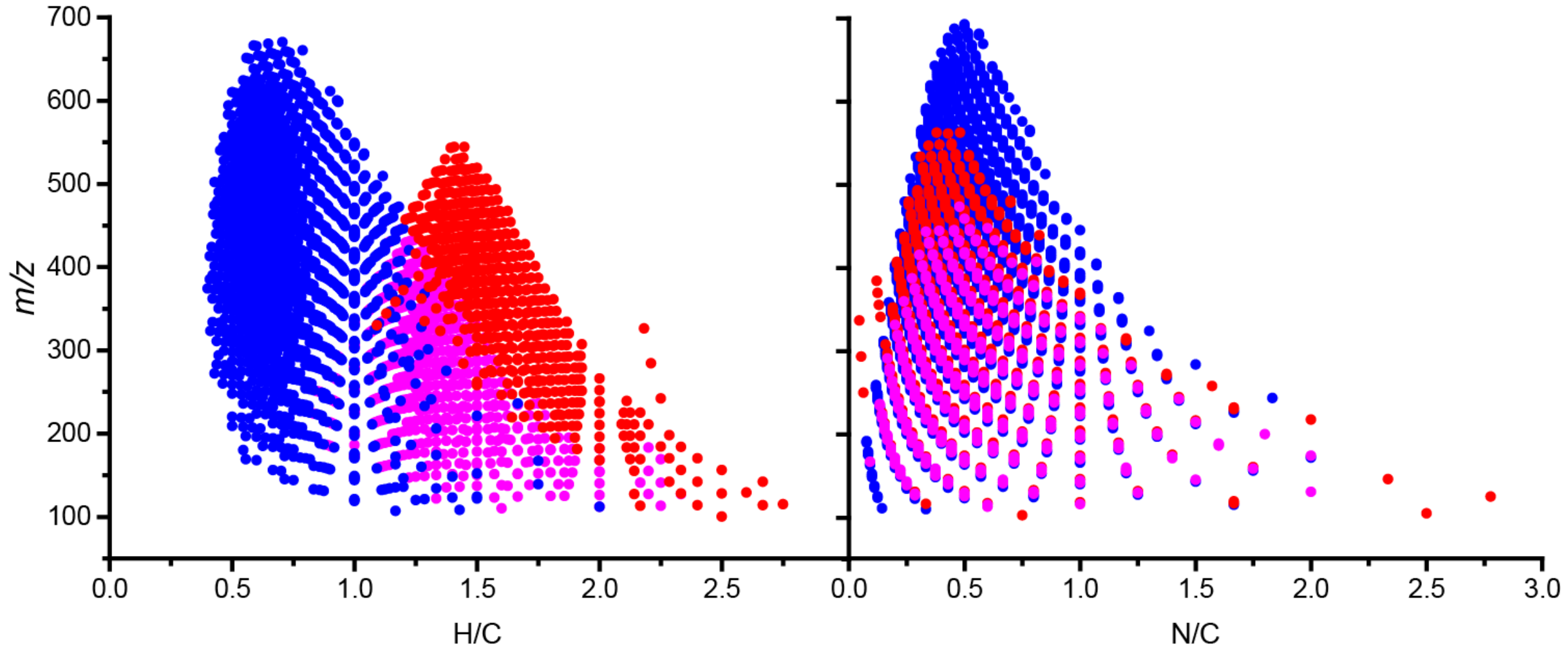

Figure 4 (2 columns): Van Krevelen diagram of the comparison between soluble and insoluble fraction with a filter at 5% of intensity. Soluble is in red, insoluble in blue and common part in purple.

Figure 4 also presents the data from the non-soluble fraction. As evidenced above, both soluble and non-soluble fractions present very different molecular structures. For the lower mass range numerous ions are present in both fractions but the difference between soluble and non-soluble fractions increases for *m/z* values larger than *m/z* 200, yielding to two distinct convergence points at high masses. The soluble fraction ions converge around H/C 1.5 and N/C 0.5 whereas the non-soluble species converge around H/C 0.75 and N/C 0.5. As discussed above, the two distribution have similar N/C ratio but they differ on the H/C ratio. The non-soluble fraction is composed of, relatively, the same amount of nitrogen as the soluble fraction, but is much less hydrogenated and so, presents a higher amount of unsaturation.

The distribution representative of the insoluble fraction tends to show that there is another set of polymeric structures in tholins that has not be identified before from the analysis of the soluble fraction only.

Previous postulate, based on IR spectroscopy, considered that soluble species became insoluble when they grew in mass and so that soluble fraction was representative of the global sample at low *m/z* value. With the results presented here, we identify a very different partition in tholins. Soluble and insoluble fractions while emerging from the same set of molecules at low *m/*z actually represent two families of compounds that are totally uncorrelated at higher *m/z* values.

**Table 2 (1 column) : Weight of each fraction according to LDI source response**

| Fraction | Average weight (%) | Convergence (H/C) |
|---|---|---|
| Soluble | 16.7 | 1.5 |
| Common | 33.3 | x |
| Insoluble | 50.0 | 0.75 |

Table 2 summarizes the relative weight of each fraction in the global sample according to LDI source response and also the convergence point of each fraction for H/C ratio. The relative weight was calculated by doing the sum of all intensities of species for each fraction. We can see that the common part represents a third of all the species, the insoluble a half and the soluble less than a quarter. This emphasizes that the global sample is mainly composed of the insoluble part, and that the soluble part, the one mainly analyzed in previous mass spectrometry study of tholins, is actually a minor component of tholins.

*Comparison between soluble and non-soluble species using DBE vs Carbon number plots*

Double Bond Equivalent (DBE) *vs* carbon number plots are generally used in petroleomic to observe the number of unsaturation in a family of species and to compare different oils (McKenna et al., 2010). In Figure 5 we adapted this graphic representation to highlight the differences between soluble species and non-soluble tholins species.

A specific range of molecules was selected to create these graphics as the use of the all data set lead to the overlap of multiple species. The Figure 5 presents the DBE vs Carbon atoms number plot for species containing between 6 and 9 nitrogen atoms for the soluble and non-soluble fractions. We added the Polycyclic Aromatic Hydrocarbon (PAH) line, which represents the most unsaturated family of species containing only carbon and hydrogen as shown previously (Cho et al., 2011). Soluble species are principally under the PAH line whereas non-soluble species are mainly above this reference line. This highlights the fact that the non-soluble species in the tholins are much more unsaturated than the soluble species. This result is also consistent with what was observed on the Van Krevelen plot. It should be pointed out that this behavior presents several analogies with asphaltenes, which correspond to the fraction of crude oil non-soluble in pentane or heptane. The molecules present in the asphaltene fraction are mainly polyaromatic, which explain the specific properties of this fraction.

Sample treatment with methanol, which efficiently solubilized polar molecules, could explain these observations: we recover in the soluble sample several families that are solubilized but which do not necessarily have the same properties. Using different steps of washing on the PAMPRE tholins sample could be, for future studies, a good approach to isolate new families of compounds among the soluble fraction.

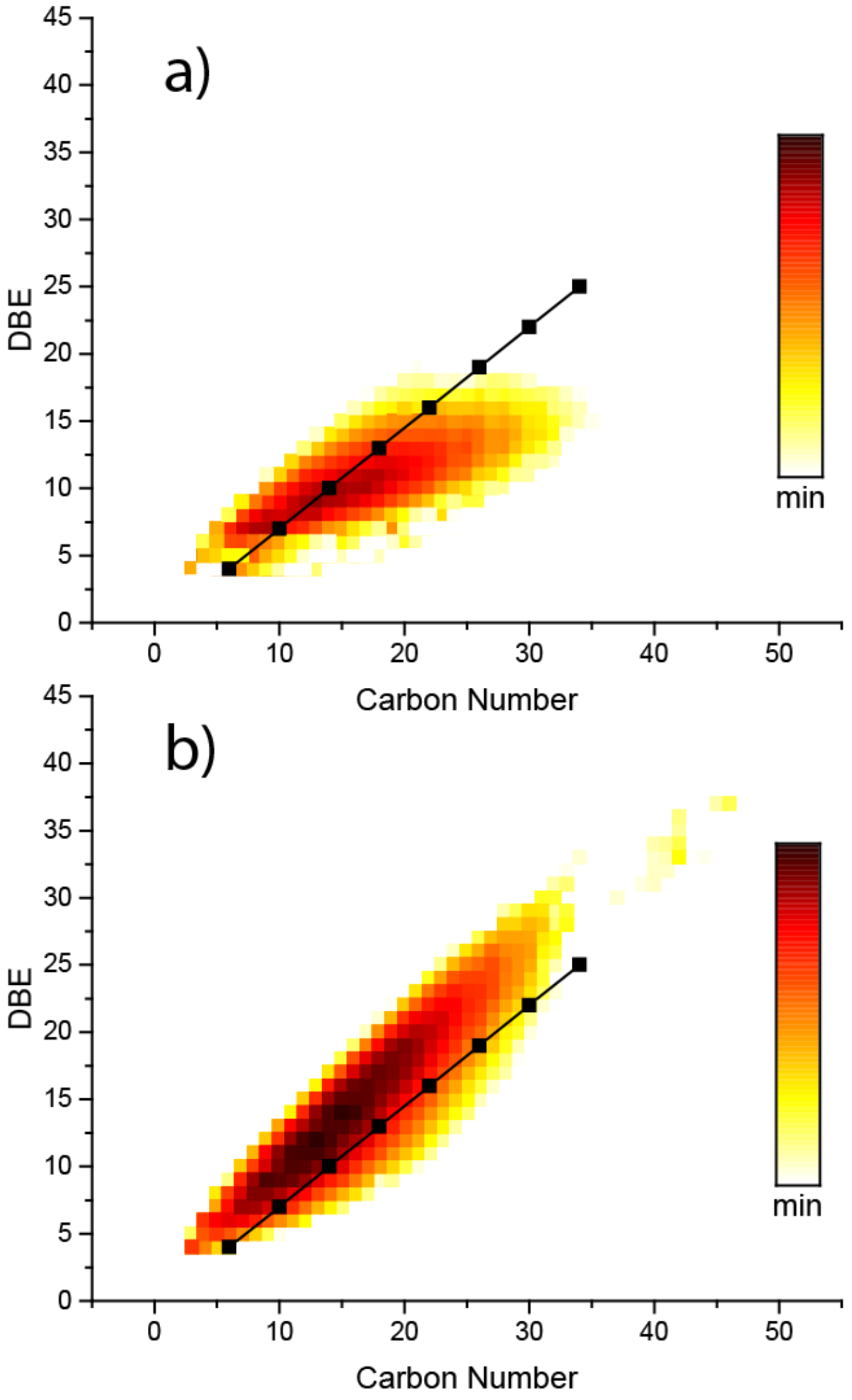

Figure 5 (1 column): a) Soluble species containing 6 to 9 Nitrogen, b) Non-soluble species containing 6 to 9 Nitrogen

## Conclusions

With our FTICR mass spectrometer equipped with a laser desorption/ionization source it was possible to analyze all the fractions of a tholins sample produced with the PAMPRE experiment (soluble in methanol, non-soluble and total) using solvent-free deposition method. The high accuracy of mass measurement coupled with the ultra-high resolution of the FTICR instrument allowed us to assign molecular formula to all species contained in the tholins. Modified Van Krevelen diagrams enabled us to easily compare the different fractions of the tholins sample. Based on infrared absorption spectroscopy signatures, it was previously postulated that soluble fraction was representative of the global sample and the only difference between soluble and insoluble species would be the *m/z* values of the molecules composing the material. In the present work, we observed significant differences between the soluble and non-soluble fractions of this sample, and showed that soluble fraction was mainly not correlated to the insoluble fraction. We highlighted that these fractions are mostly different at the molecular level when *m/z* increases and that the non-soluble fraction is much less hydrogenated than the soluble fraction. We also confirmed that the soluble fraction does not represent the major part of the sample as expected before, but the ionization yield is similar for each fraction according to their respective proportions observed with the laser ionization. Furthermore, this work demonstrated the presence of a second set of copolymeric structures within the tholins that had not been identified yet and which are less saturated than the one present in the soluble fraction.

This work highlights the importance to study of all the material and not only the soluble fraction. First, the non-soluble part represents 65% of the global mass sample and is then more representative of the bulk sample. Thus Titan's tholins have been found less hydrogenated than previously thought, based on the analysis of the soluble fraction. This high unsaturation level can be explained either by reticulation (interconnections of atoms to form network rich in cycle and single bounds), or by the presence of double and triple bonds. Yet the two structures involve different reactivities, which could impact our understanding of aerosols/gas interactions in Titan's atmosphere.

In addition, we have shown the importance of the fractionation between the soluble and insoluble part of Titan's aerosols analogues. This phenomenon could also occur at the surface of Titan at the interface between the atmosphere and the lakes and/or cryovolcanos. This would have strong implications on the organic content of Titan's lakes.

This work also confirmed the potential of the Laser Desorption Ionization source for the study of organic matter in planetary environment (Gautier et al., 2017), as being currently implemented for the analysis of Martian organic matter in the ESA/ExoMars mission (Arevalo et al., 2015). We also showed, on the case of analogs of Titan aerosol analogues, that this ionization method would be a compelling asset if coupled with high resolution mass spectrometry, as currently developed for future space missions (Briois et al., 2016).

Finally, our results on Titan's tholins can be compared to studies on the soluble (SOM) and insoluble (IOM) organic matter found in meteorites. The insoluble matter is known to be constituted of highly branched aliphatics and substituted aromatics (Derenne and Robert 2010), whereas the soluble matter is comprised of more saturated molecules (Schmitt-Koplin, 2010). The origin of both IOM and SOM and their relation with each other is highly debated. They could originate from different reservoirs. One could be the result of the processing of the other, or they could be produced by the same starting point at the beginning and then, differ when they continue to grow as we showed in this study (Derenne and Robert, 2010). Our work on a laboratory complex organic material shows that two fractions with different properties and structures can actually originate from the same formation processes.

**Acknowledgments**

N.C. thanks the European Research Council for funding via the ERC PrimChem project (grant agreement No. 636829.).

Financial support from the National FT-ICR network (FR 3624 CNRS) for conducting the research is also gratefully acknowledged.

This work was supported at COBRA laboratory by the European Regional Development Fund (ERDF) N°31708, the Région Normandie, and the Labex SynOrg (ANR-11-LABX-0029).

# References

Anicich, V.G., Wilson, P.F., McEwan, M.J., 2006. An ICR study of ion-molecules reactions relevant to Titan's atmosphere: an investigation of binary hydrocarbon mixtures up to 1 micron. Journal of the American Society for Mass Spectrometry 17, 544-561.

Arevalo, R., Brinckerhoff, W., van Amerom, F., Danell, R., Pinnick, V., Li, X., Getty, S., Hovmand, L., Grubisic, A., Mahaffy, P., Goesmann, F., Steininger, H., 2015. Design and Demonstration of the Mars Organic Molecule Analyzer (MOMA) on the ExoMars 2018 Rover. IEEE Aerospace Conference.

Barrere, C., Hubert-Roux, M., Lange, C.M., Rejaibi, M., Kebir, N., Desilles, N., Lecamp, L., Burel, F., Loutelier-Bourhis, C., 2012. Solvent-based and solvent-free characterization of low solubility and low molecular weight polyamides by mass spectrometry: a complementary approach. Rapid communications in mass spectrometry : RCM 26, 1347-1354.

Bonnet, J.-Y., Thissen, R., Frisari, M., Vuitton, V., Quirico, É., Orthous-Daunay, F.-R., Dutuit, O., Le Roy, L., Fray, N., Cottin, H., Hörst, S.M., Yelle, R.V., 2013. Compositional and structural investigation of HCN polymer through high resolution mass spectrometry. International Journal of Mass Spectrometry 354-355, 193-203.

Briois, C., Thissen, R., Thirkell, L., Aradj, K., Bouabdellah, A., Boukrara, A., Carrasco, N., Chalumeau, G., Chapelon, O., Colin, F., Coll, P., Cottin, H., Engrand, C., Grand, N., Lebreton, J.-P., Orthous-Daunay, F.-R., Pennanech, C., Szopa, C., Vuitton, V., Zapf, P., Makarov, A., 2016. Orbitrap mass analyser for in situ characterisation of planetary environments: Performance evaluation of a laboratory prototype. Planetary and Space Science 131, 33-45.

Cable, M.L., Hörst, S.M., He, C., Stockton, A.M., Mora, M.F., Tolbert, M.A., Smith, M.A., Willis, P.A., 2014. Identification of primary amines in Titan tholins using microchip nonaqueous capillary electrophoresis. Earth and Planetary Science Letters 403, 99-107.

Cable, M.L., Horst, S.M., Hodyss, R., Beauchamp, P.M., Smith, M.A., Willis, P.A., 2012. Titan tholins: simulating Titan organic chemistry in the Cassini-Huygens era. Chemical reviews 112, 1882-1909.

Carrasco, N., Jomard, F., Vigneron, J., Etcheberry, A., Cernogora, G., 2016. Laboratory analogues simulating Titan's atmospheric aerosols: Compared chemical compositions of grains and thin films. Planetary and Space Science 128, 52-57.

Carrasco, N., Schmitz-Afonso, I., Bonnet, J.Y., Quirico, E., Thissen, R., Dutuit, O., Bagag, A., Laprevote, O., Buch, A., Giulani, A., Adande, G., Ouni, F., Hadamcik, E., Szopa, C., Cernogora, G., 2009. Chemical characterization of Titan's tholins: solubility, morphology and molecular structure revisited. The journal of physical chemistry. A 113, 11195-11203.

Cho, Y., Kim, Y.H., Kim, S., 2011. Planar limit-assisted structural interpretation of saturates/aromatics/resins/asphaltenes fractionated crude oil compounds observed by Fourier transform ion cyclotron resonance mass spectrometry. Analytical chemistry 83, 6068-6073.

Coll, P., Coscia, D., Smith, N., Gazeau, M.-C., Ramirez, S.I., Cernogora, C., Israël, G., Raulin, F., 1999. Experimantal laboratory simulation of Titan's atmosphere: aerosols and gas phase. Planetary and Space Science 47, 1331-1340.

Derenne, S., Robert, F., 2010. Model of molecular structure of the insoluble organic matter isolated from Murchison meteorite. Meteoritics & Planetary Science 45, 1461-1475.

Fulchignoni, M., Ferri, F., Angrilli, F., Ball, A.J., Bar-Nun, A., Barucci, M.A., Bettanini, C., Bianchini, G., Borucki, W., Colombatti, G., Coradini, M., Coustenis, A., Debei, S., Falkner, P., Fanti, G., Flamini, E., Gaborit, V., Grard, R., Hamelin, M., Harri, A.M., Hathi, B., Jernej, I., Leese, M.R., Lehto, A., Lion Stoppato, P.F., Lopez-Moreno, J.J., Makinen, T., McDonnell, J.A., McKay, C.P., Molina-Cuberos, G., Neubauer, F.M., Pirronello, V., Rodrigo, R., Saggin, B., Schwingenschuh, K., Seiff, A., Simoes, F., Svedhem, H., Tokano, T., Towner, M.C., Trautner, R., Withers, P., Zarnecki, J.C., 2005. In situ measurements of the physical characteristics of Titan's environment. Nature 438, 785-791.

Gautier, T., Carrasco, N., Buch, A., Szopa, C., Sciamma-O'Brien, E., Cernogora, G., 2011. Nitrile gas chemistry in Titan's atmosphere. Icarus 213, 625-635.

Gautier, T., Carrasco, N., Mahjoub, A., Vinatier, S., Giuliani, A., Szopa, C., Anderson, C.M., Correia, J.-J., Dumas, P., Cernogora, G., 2012. Mid- and far-infrared absorption spectroscopy of Titan's aerosols analogues. Icarus 221, 320-327.
Gautier, T., Carrasco, N., Schmitz-Afonso, I., Touboul, D., Szopa, C., Buch, A., Pernot, P., 2014. Nitrogen incorporation in Titan's tholins inferred by high resolution orbitrap mass spectrometry and gas chromatography–mass spectrometry. Earth and Planetary Science Letters 404, 33-42.
Gautier, T., Schmitz-Afonso, I., Touboul, D., Szopa, C., Buch, A., Carrasco, N., 2016. Development of HPLC-Orbitrap method for identification of N-bearing molecules in complex organic material relevant to planetary environments. Icarus 275, 259-266.
Gautier, T., Sebree, J.A., Li, X., Pinnick, V.T., Grubisic, A., Loeffler, M.J., Getty, S.A., Trainer, M.G., Brinckerhoff, W.B., 2017. Influence of trace aromatics on the chemical growth mechanisms of Titan aerosol analogues. Planetary and Space Science 140, 27-34.
He, C., Lin, G., Smith, M.A., 2012. NMR identification of hexamethylenetetramine and its precursor in Titan tholins: Implications for Titan prebiotic chemistry. Icarus 220, 627-634.
Hörst, S.M., Yoon, Y.H., Ugelow, M.S., Parker, A.H., Li, R., de Gouw, J.A., Tolbert, M.A., 2018. Laboratory investigations of Titan haze formation: In situ measurement of gas and particle composition. Icarus 301, 136-151.
Hughey, C.A., Hendrickson, C.L., Rodgers, R.P., Marshall, A.G., 2001. Kendrick Mass Defect Spectrum: A Compact Visual Analysis for Ultrahigh-Resolution Broadband Mass Spectra. Anal. Chem. 73, 4676-4681.
Imanaka, H., Khare, B.N., Elsila, J.E., Bakes, E.L.O., McKay, C.P., Cruikshank, D.P., Sugita, S., Matsui, T., Zare, R.N., 2004. Laboratory experiments of Titan tholin formed in cold plasma at various pressures: implications for nitrogen-containing polycyclic aromatic compounds in Titan haze. Icarus 168, 344-366.
Imanaka, H., Smith, M.A., 2010. Formation of nitrogenated organic aerosols in the Titan upper atmosphere. Proceedings of the National Academy of Sciences 107.
Israel, G., Szopa, C., Raulin, F., Cabane, M., Niemann, H.B., Atreya, S.K., Bauer, S.J., Brun, J.F., Chassefiere, E., Coll, P., Conde, E., Coscia, D., Hauchecorne, A., Millian, P., Nguyen, M.J., Owen, T., Riedler, W., Samuelson, R.E., Siguier, J.M., Steller, M., Sternberg, R., Vidal-Madjar, C., 2005. Complex organic matter in Titan's atmospheric aerosols from in situ pyrolysis and analysis. Nature 438, 796-799.
Kendrick, E., 1963. A Mass Scale Based on CH2= 14.0000 for High Resolution Mass Spectrometry of Organic Compounds. Analytical chemistry 35, 2146-2154.
Kim, S., Kramer, R.W., Hatcher, P.G., 2003. Graphical Method for Analysis of Ultrahigh-Resolution Broadband Mass Spectra of Natural Organic Matter, the Van Krevelen Diagram. Anal. Chem. 75, 5336-5344.
Mahjoub, A., Schwell, M., Carrasco, N., Benilan, Y., Cernogora, G., Szopa, C., Gazeau, M.-C., 2016. Characterization of aromaticity in analogues of titan's atmospheric aerosols with two-step laser desorption ionization mass spectrometry. Planetary and Space Science 131, 1-13.
McKay, C.P., 1996. Elemental composition, solubility, and optical properties of Titan's organic haze. Pergamon 44.
McKenna, A.M., Purcell, J.M., Rodgers, R.P., Marshall, A.G., 2010. Heavy Petroleum Composition. 1. Exhaustive Compositional Analysis of Athabasca Bitumen HVGO Distillates by Fourier Transform Ion Cyclotron Resonance Mass Spectrometry: A Definitive Test of the Boduszynski Model. Energy & Fuels 24, 2929-2938.
Neish, C.D., Somogyi, Á., Lunine, J.I., Smith, M.A., 2009. Low temperature hydrolysis of laboratory tholins in ammonia-water solutions: Implications for prebiotic chemistry on Titan. Icarus 201, 412-421.
Niemann, H.B., Atreya, S.K., Bauer, S.J., Carignan, G.R., Demick, J.E., Frost, R.L., Gautier, D., Haberman, J.A., Harpold, D.N., Hunten, D.M., Israel, G., Lunine, J.I., Kasprzak, W.T., Owen, T.C., Paulkovich, M., Raulin, F., Raaen, E., Way, S.H., 2005. The abundances of constituents of Titan's atmosphere from the GCMS instrument on the Huygens probe. Nature 438, 779-784.

Pereira, T.M.C., Vanini, G., Tose, L.V., Cardoso, F.M.R., Fleming, F.P., Rosa, P.T.V., Thompson, C.J., Castro, E.V.R., Vaz, B.G., Romão, W., 2014. FT-ICR MS analysis of asphaltenes: Asphaltenes go in, fullerenes come out. Fuel 131, 49-58.
Pernot, P., Carrasco, N., Thissen, R., Schmitz-Afonso, I., 2010. Tholinomics—Chemical Analysis of Nitrogen-Rich Polymers. Analytical chemistry 82, 1371-1380.
Poch, O., Coll, P., Buch, A., Ramírez, S.I., Raulin, F., 2012. Production yields of organics of astrobiological interest from H2O–NH3 hydrolysis of Titan's tholins. Planetary and Space Science 61, 114-123.
Sagan, C., Khare, B.N., Thompson, W.R., McDonald, G.D., Wing, M.R., Bada, J.L., Vo-Dinh, T., Arakawa, E.T., 1993. Polycyclic aromatic hydrocarbons in the atmospheres of Titan and Jupiter. The Astrophysical Journal 414, 399.
Sagan, C., Thompson, W.R., Khare, B.N., 2002. Titan: a laboratory for prebiological organic chemistry. Accounts of Chemical Research 25, 286-292.
Sciamma-O'Brien, E., Upton, K.T., Salama, F., 2017. The Titan Haze Simulation (THS) experiment on COSmIC. Part II. Ex-situ analysis of aerosols produced at low temperature. Icarus 289, 214-226.
Somogyi, A., Oh, C.H., Smith, M.A., Lunine, J.I., 2005. Organic environments on Saturn's moon, Titan: simulating chemical reactions and analyzing products by FT-ICR and ion-trap mass spectrometry. Journal of the American Society for Mass Spectrometry 16, 850-859.
Somogyi, Á., Smith, M.A., Vuitton, V., Thissen, R., Komáromi, I., 2012. Chemical ionization in the atmosphere? A model study on negatively charged "exotic" ions generated from Titan's tholins by ultrahigh resolution MS and MS/MS. International Journal of Mass Spectrometry 316-318, 157-163.
Swaraj, S., Oran, U., Lippitz, A., Friedrich, J.F., Unger, W.E.S., 2007. Aging of Plasma-Deposited Films Prepared from Organic Monomers. Plasma Processes and Polymers 4, S784-S789.
Szopa, C., Cernogora, G., Boufendi, L., Correia, J.J., Coll, P., 2006. PAMPRE: A dusty plasma experiment for Titan's tholins production and study. Planetary and Space Science 54, 394-404.
Tanaka, R., Sato, S., Takanohashi, T., Hunt, J.E., Winans, R.E., 2004. Analysis of the Molecular Weight Distribution of Petroleum Asphaltenes Using Laser Desorption-Mass Spectrometry. Energy & Fuels 18, 1405-1413.
Toupance, G., Raulin, F., Buvet, R., 1975. Formation of prebiochemical compounds in models of the primitive Earth's atmosphere. Origins of Life 6, 83-90.
Tran, B., Ferris, J., Chera, J., 2003. The photochemical formation of a titan haze analog. Structural analysis by x-ray photoelectron and infrared spectroscopy. Icarus 162, 114-124.
Van Krevelen, D.W., 1950. Graphical-statistical method for the study of structure and reaction processes of coal. Fuel 29, 269-284.
Vuitton, V., Bonnet, J.-Y., Frisari, M., Thissen, R., Quirico, E., Dutuit, O., Schmitt, B., Le Roy, L., Fray, N., Cottin, H., Sciamma-O'Brien, E., Carrasco, N., Szopa, C., 2010. Very high resolution mass spectrometry of HCN polymers and tholins. Faraday Discussions 147, 495.

**Supporting information for:**

*Comparison between different ionization sources and instruments on tholins 5%*

As we can see on Figure S1, a study was made to validate our results obtained with laser desorption ionization method.

First, we compared spectra recorded on a LDI-TOF instrument and LDI-FTICR instrument to observe if different analyzer equipped with the same ionization source gave the same results. Except for the resolution, no differences were founded on spectra. Figures S1a and S1b): the general pattern of the mass spectrum is similar except for higher intensities for lower *m/z* values. This is due to transmission parameters of the SolariX mass spectrometer which tends to be more discriminating than the TOF analyzer. Nevertheless, we confirm that we detect the same species with the same repetition patterns with both analyzers and therefore we validate the species detected by FTICR.

We, then, compared two different ionization sources on the same analyzer (Figure S1b, S1c). We recovered quite the same distribution, with a base peak located at a low *m/z* value.

Figure S2 shows a zoom on *m/z* 155 and *m/z* 504. Same species were recovered for each ionization method and this allowed us to validate our results.

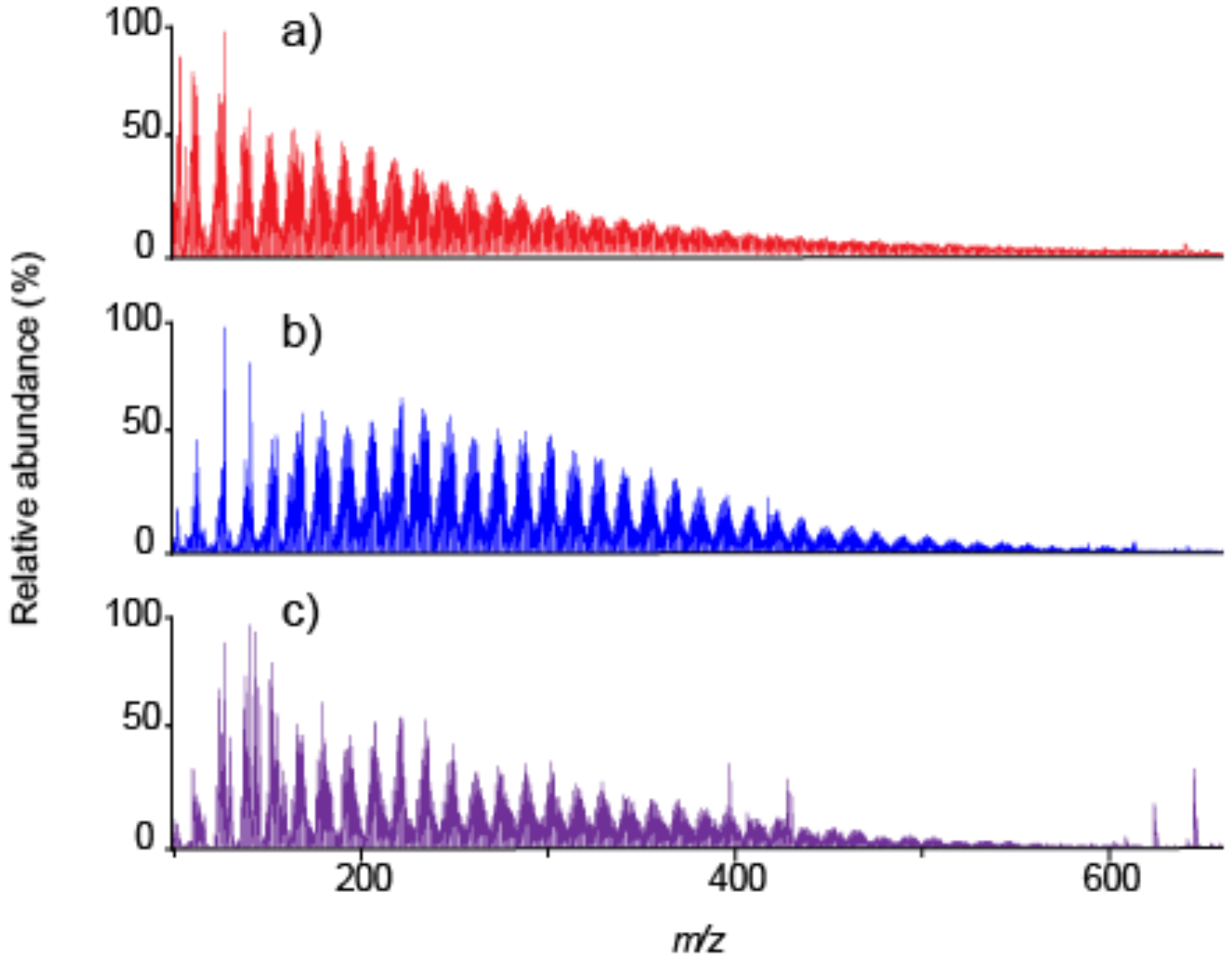

Figure S1: Comparison between a) LDI-TOF b) LDI-FTICR and c) ESI-FTICR of soluble fraction of tholins 5%

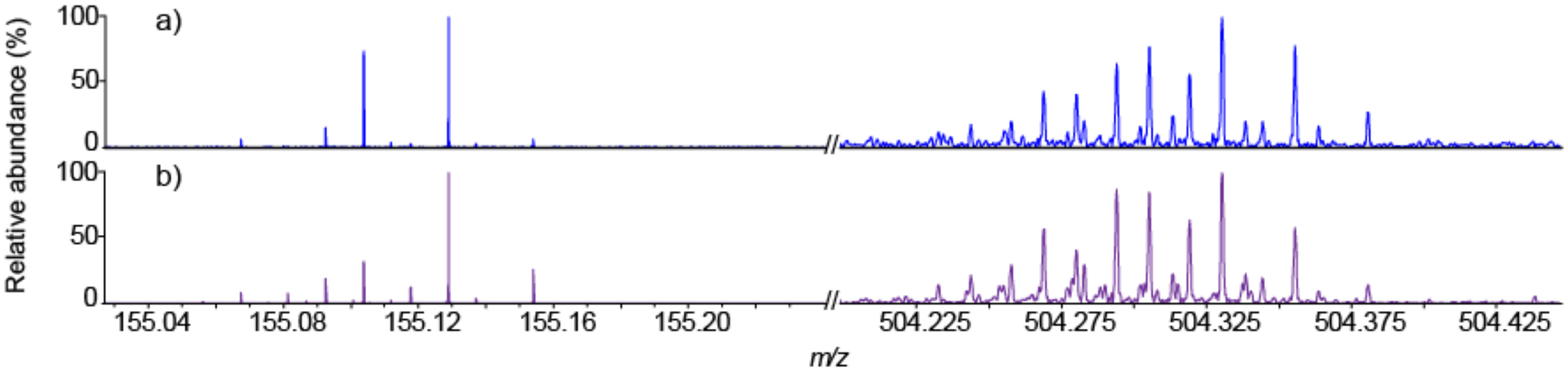

Figure S2: Comparison at *m/z* 155 and *m/z* 504 for the a) LDI-FTICR and b) ESI-FTICR soluble fraction of tholins 5%

*Threshold*

Because fractions responses to the laser ionization are different due to their composition (one is quite more saturated than this other), it was impossible to compare them with the same laser power. In fact, with the same ionization energy, spectra were not usable. Insoluble fraction was overexpressed and the peak shape was not good and soluble fraction got a global intensity which was very low. We decided, to avoid this problem, to observe the variation of the global intensity with the rise of the laser power for each fraction. Figure 3 shows the result of this comparison after the normalization of each sum of intensities. We observe that each fraction got quite the same curve shape.

To accommodate the different behaviour of soluble and non-soluble fractions, and the global sample, we have to define the best laser power to be used for each sample. The laser delivers a maximum power of 166.7 kJ per shot. We first determined the minimum power to be applied to detect the first ions, 14%Pmax for the soluble and 24%Pmax for the insoluble. The best compromise between a good signal and avoiding ion coalescence, was found by settling the laser power at 5% above the threshold power, i.e. 19%Pmax and 29%Pmax for respectively the insoluble and soluble fractions. The global intensity was for all samples 1.10e7 for 10 scans.

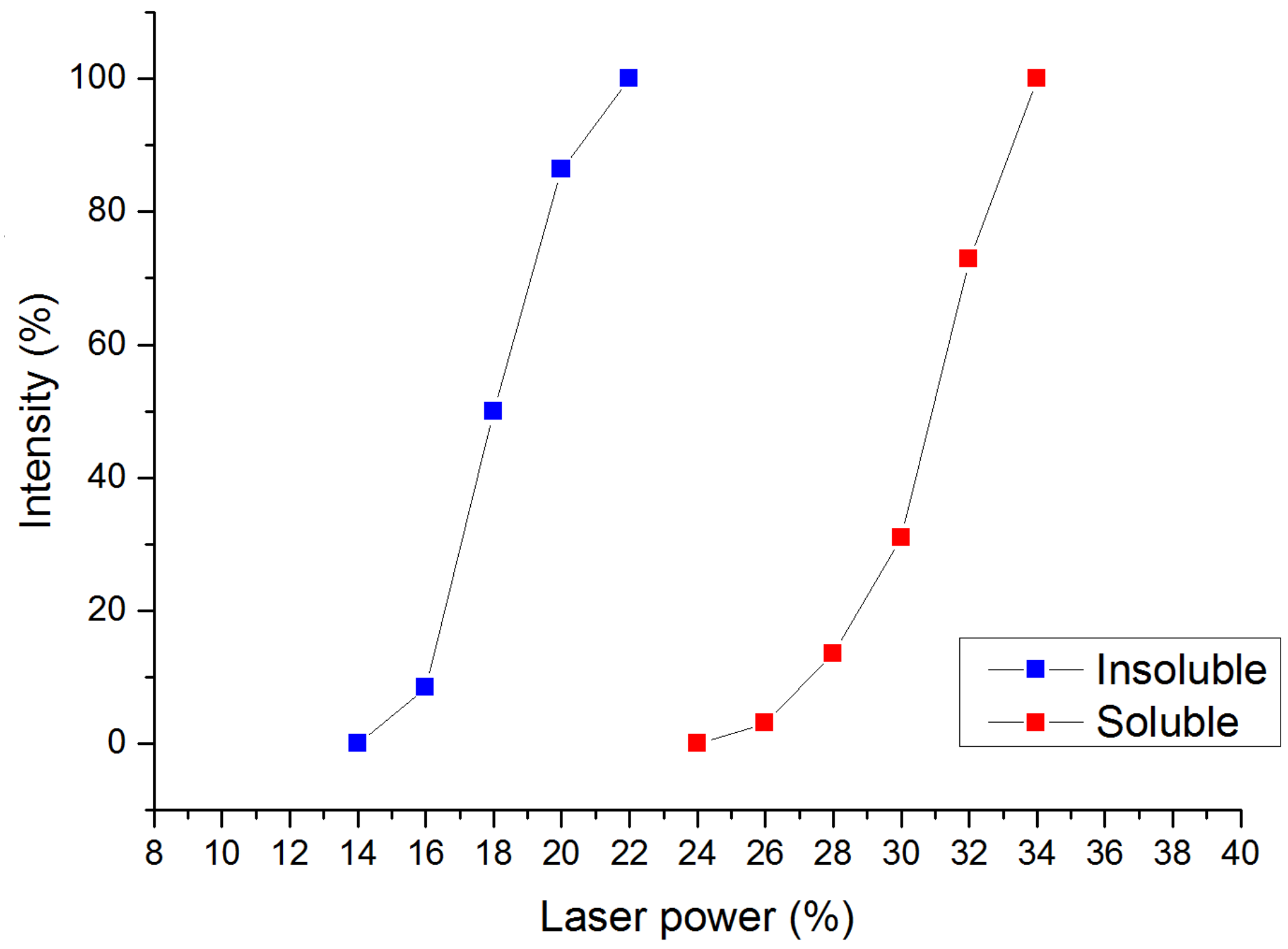

Figure S3: Evolution of the signal intensity in LDI as a function of the laser power for the non-soluble in blue and soluble in red tholins fractions.

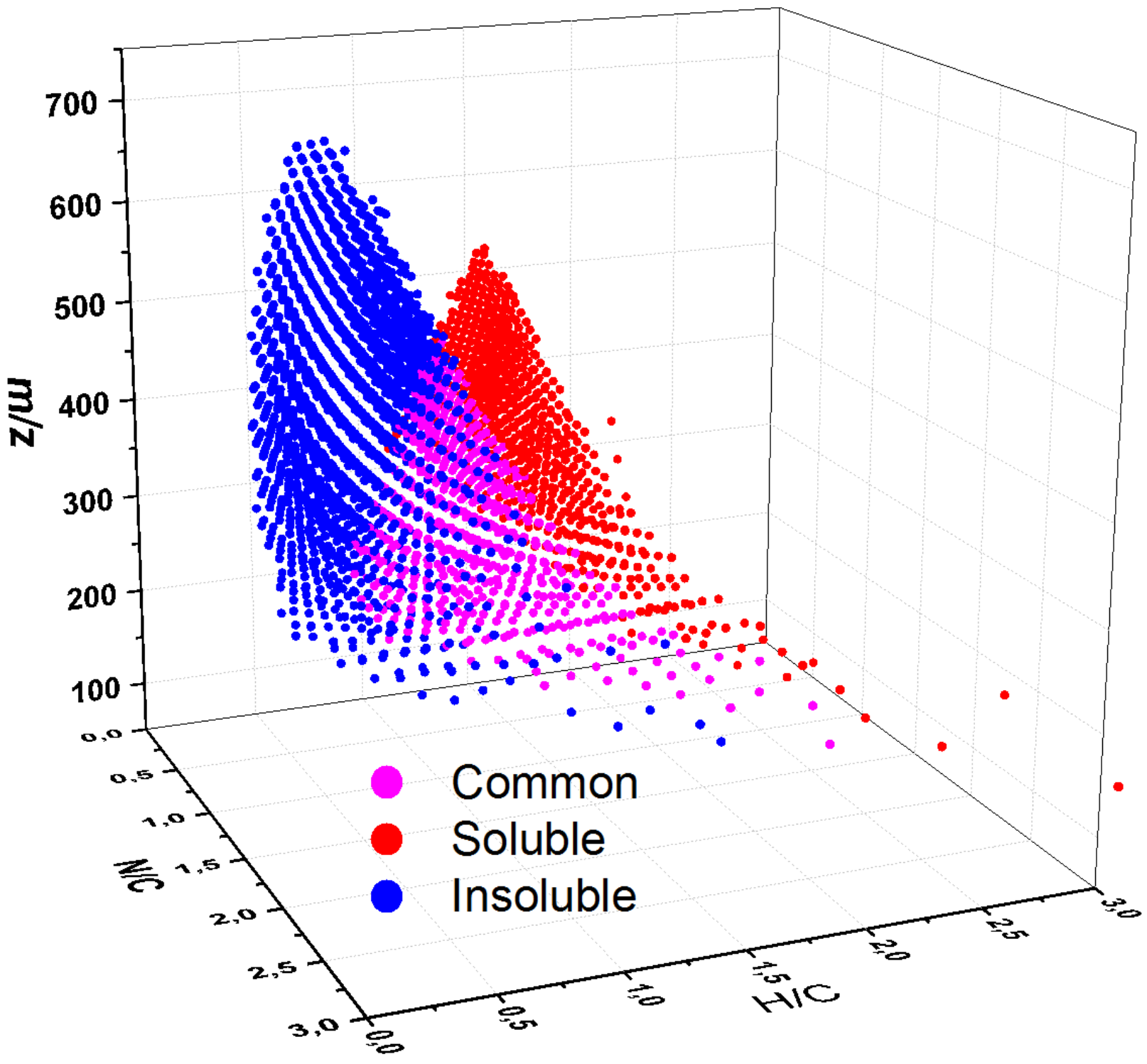

Figure S4: 3D Van Krevelen for the comparison of tholins fractions with a filter at 5% of intensity. Soluble is in red, insoluble in blue and common part in purple.